\documentstyle[aps,preprint]{revtex}

\begin{document}

\title{Mesoscopic non-equilibrium thermodynamics approach to the dynamics of polymers \footnote{We dedicate this paper to Dick Bedeaux in the occasion of his 60th birthday}}

\author{J.M. Rub\'{\i} and A. P\'erez-Madrid}
\address{Departament de F\'{\i}sica Fonamental and CER F\'{\i}sica de Sistemes Complexos,\\
Facultat de F\'{\i}sica,\\
Universitat de Barcelona,\\
Diagonal 647, 08028 Barcelona, Spain\\}

\maketitle

\begin{abstract}
We present a general formalism able to derive the kinetic equations of polymer dynamics. It is based on the application of nonequilibrium thermodynamics to analyze the irreversible processes taking place in the conformational space of the macromolecules. The Smoluchowski equation results from the analysis of the underlying diffusion process in that space within the scheme of nonequilibrium thermodynamics. We apply the method to different situations, concerning flexible, semiflexible and rod-like polymers and to the case of more concentrated solutions in which interactions become important.
\end{abstract}

\section{Introduction}

In the theory of polymer dynamics \cite{kn:doi}, \cite{kn:yamakawa}, and in general in the dynamics of complex fluids, the knowledge of the kinetic equations describing the evolution of the probability density for the configurations of the suspended phase plays a central role. The formulation of  Fokker-Planck or Smoluchowski equations constitutes the starting point in the study of the configurations of the system and in the characterization of its global behaviour. It is then of primary importance to establish methods able to provide those equations for a complex system possessing a huge number of interacting degrees of freedom and exhibiting a wide variety of time and length scales.

>From the statistical mechanical point of view, a polymer solution can be conceived as a set of objects of mesoscopic size immersed in a heat bath. The different configurations adopted by the macromolecules depend on their intrinsic nature, the interactions between their different parts and on the interactions with the bath. All these ingredients must be considered in the formulation of Fokker-Planck equations.

Our purpose in this paper is to present a general and simple formalism able to derive kinetic equations of the Fokker-Planck type for polymers in solution. The formalism is based on the fact that processes leading to variations in the conformation of the macromolecules can be described by nonequilibrium thermodynamics. The extension of that theory to the mesoscopic level of description, referred to as mesoscopic nonequilibrium thermodynamics (MNET), has been applied to different situations pertaining to the domain of transport phenomena \cite{kn:mazur}-\cite{kn:ivan} and activated processes \cite{kn:agusti2}, \cite{kn:david}. We will show that it can also be used in the case of polymers to systematically derive the Smoluchowski equations for polymers of different nature.

The paper is organized in the following way. In section 2, we present the general derivation of the kinetic equation for a macromolecule in solution in the framework of MNET. These results are applied in Section 3 to the case of flexible and rod-like polymers to derive the corresponding Smoluchowski equations. The case of semiflexible polymers is presented in section 4. We derive the functional Fokker-Planck equation for the probability of the different configurations of the chain. In Section 5, we treat the case of more concentrated solutions in which interactions between segments belonging to different chains become important. The functional Fokker-Planck equation is obtained in terms of the concentration of segments. Finally, in Section 6 we present our main conclusions. 

\section{Kinetic equations for the dynamics of macromolecules in solution}

A polymer solution is usually modeled as a suspension of macromolecules in a Newtonian solvent \cite{kn:doi}, \cite{kn:degennes}. This simplification may be carried out due to the existence of well-separated time and length scales related to the carrier fluid and to the macromolecules. The dynamics of the fluid itself is then governed, at the continuum level, by the Navier-Stokes equations whereas the polymers are considered as suspended objects whose presence has implications in the dynamics of the whole system. Since the interaction between the polymer and the solvent occurs through its contour, the form of the polymers will be of crucial importance in the characterization of the properties of the system.

The polymer solution is usually conceived as a two-component system. One of them, is the host fluid that plays the role of a heat bath at constant temperature T. To characterize the different configurations of the macromolecule, we will introduce the vector $\underline{\gamma}$ whose components constitute the independent parameters necessary to completely determine its configuration. We can then define the quantity $P(\underline{\gamma},t)/N$ as the probability density for the polymer in the configuration space ($\underline{\gamma}$-space), where $N$ is the number of polymers in suspension.

Since $P(\underline{\gamma},t)$ is a conserved quantity, it evolves according to the continuity equation

\begin{equation}\label{eq:a2}
\frac{\partial P(\underline{\gamma},t)}{\partial t} = -\nabla_{\underline{\gamma}}\cdot\underline{J}(\underline{\gamma},t)
\end{equation}

\noindent with $\underline{J}(\underline{\gamma},t)$ being a current in the $\underline{\gamma}$-space.

Non-equilibrium thermodynamics establishes the Gibbs equation in which changes in the entropy are related to variations in the density of the particles. We will generalize that idea by assuming that variations in the probability density in $\underline{\gamma}$-space are responsible for changes in the total entropy of the system, S. Our starting point will then be to formulate the Gibbs equation

\begin{equation}\label{eq:a4}
\delta S = -\frac{1}{T}\int \mu(\underline{\gamma},t) \delta P(\underline{\gamma},t) d\underline{\gamma} \; ,
\end{equation}

\noindent where $\mu(\underline{\gamma},t)$ is a chemical potential defined in $\underline{\gamma}$-space. This equation holds for constant energy and volume of the system. 

The expression of the chemical potential can be identified by imposing that the Gibbs equation (\ref{eq:a4}) must be compatible with the Gibbs entropy postulate. This postulate establishes that 

\begin{equation}\label{eq:a5}
S = -k \int P \ln P/P^{l.eq.} d\underline{\gamma} + S^{l.eq.} \; ,
\end{equation}

\noindent where $k$ is the Boltzmann constant, and the local equilibrium distribution $P^{l.eq.}$ is 

\begin{equation}\label{eq:a6}
P^{l.eq.}(\underline{\gamma},t) = \exp\{\left(N\mu_p - U\right)/kT\}\; ,
\end{equation}

\noindent with $\mu_p$ being the chemical potential of the polymers at local equilibrium, and $U$ the potential energy. In addition, $S^{l.eq.}$ is the total entropy at local equilibrium whose variations are given by the Gibbs equation

\begin{equation}\label{eq:a7}
\delta S^{l.eq.} = - \frac{1}{T} \mu_p \delta N\; .
\end{equation}

\noindent By taking variations in Eq. (\ref{eq:a5}) and comparing with Eq. (\ref{eq:a4}), after using Eq. (\ref{eq:a7}) we obtain the expression for the chemical potential $\mu(\underline{\gamma},t)$ 

\begin{equation}\label{eq:a8}
\mu(\underline{\gamma},t) = \mu_p + \frac{kT}{N} \ln (P/P^{l.eq.}) = \frac{kT}{N} \ln P + \frac{U}{N}\; .
\end{equation}

\noindent The Gibbs equation then resembles its corresponding expression for an ideal mixture in which the different species would correspond to the different values of $\underline{\gamma}$, which may be interpreted as an internal coordinate or degree of freedom \cite{kn:mazur}. 

The entropy production, accounting for dissipation in the  $\underline{\gamma}$-space related to variations of the configuration of the polymer follows straightforwardly from the Gibbs equation (\ref{eq:a4}). Taking the time derivative in Eq. (\ref{eq:a4}) and using Eq. (\ref{eq:a2}), after partial integration we obtain 

\begin{equation}\label{eq:a9}
\sigma = -\frac{1}{T}\int \underline{J}(\underline{\gamma},t)\cdot\nabla_{\underline{\gamma}}\mu(\underline{\gamma},t) d\underline{\gamma}\; .
\end{equation}

\noindent Following the rules of nonequilibrium thermodynamics \cite{kn:mazur}, from the entropy production one derives the linear laws 
\begin{equation}\label{eq:a10}
\underline{J}(\underline{\gamma},t) = -\frac{1}{T}\int\underline{L}(\underline{\gamma},\underline{\gamma}\prime)\nabla_{\underline{\gamma}\prime}\mu(\underline{\gamma}\prime,t) d\underline{\gamma}\prime\; ,
\end{equation}

\noindent where $\underline{L}(\underline{\gamma},\underline{\gamma}\prime)$ is a matrix of phenomenological coefficients. Its local version, occurring when $\underline{L}(\underline{\gamma},\underline{\gamma}\prime) = \underline{L}(\underline{\gamma})\delta(\underline{\gamma}-\underline{\gamma}\prime)$, is then given by 

\begin{equation}\label{eq:a11}
\underline{J}(\underline{\gamma},t) = -\frac{1}{T}\underline{L}(\underline{\gamma})\nabla_{\underline{\gamma}}\mu(\underline{\gamma},t)\; .
\end{equation}

\noindent Inserting the expression of the chemical potential Eq. (\ref{eq:a8}) into Eq. (\ref{eq:a11}) and the resulting current into the continuity equation (\ref{eq:a2}) one obtains

\begin{equation}\label{eq:a12}
\frac{\partial P}{\partial t} = \frac{1}{T}\nabla_{\underline{\gamma}}\cdot\underline{L}\cdot\nabla_{\underline{\gamma}}\left(\frac{kT}{N}\ln P + U\right)
\end{equation}

\noindent or

\begin{equation}\label{eq:a13}
\frac{\partial P}{\partial t} = \nabla_{\underline{\gamma}}\cdot\underline{M}\cdot (kT\nabla_{\underline{\gamma}}P + P\nabla_{\underline{\gamma}}U)\; ,
\end{equation}

\noindent where use has been made of the relation between the matrices of phenomenological coefficients and the matrices $\underline{M}$

\begin{equation}\label{eq:a14}
\underline{M} = \frac{\underline{L}}{NTP}\; .
\end{equation}

\noindent If we interpret $\underline{M}$ as the mobility matrix related to the dynamics of the vector $\underline{\gamma}$, Eq. (\ref{eq:a13}) is precisely the Smoluchowski equation describing the evolution of the probability density in $\underline{\gamma}$-space. To solve the Smoluchowski equation, we need to know the expression for the mobilities. This quantities can be borrowed from hydrodynamics \cite{kn:bonet}, \cite{kn:bonet1}.

\section{Flexible and rod-like polymers}

As a first example, we will apply the results of the previous section to study the dynamics of a dilute solution of flexible polymers. The macromolecules can be modeled as a set of $n$ segments (monomers) whose positions will be denoted by $\underline{\gamma} \equiv (\vec{R}_1, \vec{R}_2, ...., \vec{R}_n)$ \cite{kn:doi}. The quantity $\underline{\gamma}$ is then considered as the internal coordinate and its different values are related to a point in a 3n-dimensional configurational space. We can now introduce the density $P(\underline{\gamma},t)$ which evolves according to the continuity equation

\begin{equation}\label{eq:b1}
\frac{\partial P}{\partial t} = -\sum_{i=1}^n\nabla_{\vec{R}_i}\cdot\left(\vec{J}_i + P\vec{v}_i\right)\; ,
\end{equation}

\noindent where $\vec{J}_i$ is the current associated with the diffusion of the probability density along the coordinate $\vec{R}_i$, and $\vec{v}_i(\vec{R}_i,t) = \vec{v}_0(\vec{r}=\vec{R}_i,t)$, with $\vec{v}_0$ being an externally imposed velocity field. 

We can also define the Gibbs equation from Eq. (\ref{eq:a4}) for the polymer phase. As we did in the previous section, from that equation, and using Eq. (\ref{eq:b1}) it is possible to compute the entropy production related to the diffusion process in the conformational space

\begin{equation}\label{eq:b4} 
\sigma = -\frac{1}{T}\int \sum_{i=1}^{n}\vec{J}_i\cdot\nabla_{\vec{R}_i}\mu \; d\{\vec{R}_i\}\; .
\end{equation}

\noindent From this quantity one then derives the linear laws

\begin{equation}\label{eq:b5}
\vec{J}_i = -\frac{1}{T}\sum_{j=1}^n\underline{L}_{ij}\cdot\nabla_{\vec{R}_i}\mu\; ,
\end{equation}

\noindent where $\underline{L}_{ij}$ are now matrices of phenomenological coefficients which according to the Onsager's reciprocity principle satisfy the relations

\begin{equation}\label{eq:b6}
\underline{L}_{ij} = \underline{L}_{ji}^\dagger\; .
\end{equation}

\noindent Here the superscript $\dagger$ stands for the Hermitian adjoint. The chemical potential is also given by Eq. (\ref{eq:a8}), where now the potential energy $U(\underline{\gamma},t)$ may in general include contributions due to intermolecular interactions among monomers, external forces or even contributions coming from the presence of obstacles. The Smoluchowski equation then follows from Eqs. (\ref{eq:b1}), (\ref{eq:b5}), and (\ref{eq:a8}):

\begin{equation}\label{eq:b10}
\frac{\partial P}{\partial t} = \sum_{ij}\nabla_{\vec{R}_i}\cdot\underline{M}_{ij}\cdot\left(kT\nabla_{\vec{R}_j}P + P\nabla_{\vec{R}_j}U\right)- \sum_{i=1}^n\nabla_{\vec{R}_i}\cdot\left(P\vec{v }_i\right)\; .
\end{equation}

As a second example we will consider the case of a rod-like polymer \cite{kn:doi} which will be modeled as a cylinder of length $L$ and diameter $b$. The internal coordinate is now taken as $\underline{\gamma} = (\vec{R},\vec{n})$, with $\vec{R}$ being the centre of mass position vector and $\vec{n}$ the unit vector pointing along the axis of the polymer.

The continuity equation for   $P(\vec{R},\vec{n},t)$ is now written as 

\begin{equation}\label{eq:b11}
\frac{\partial P}{\partial t} = - \nabla_{\vec{R}}\cdot\vec{J}_{\vec{R}} - \nabla_{\vec{n}}\cdot\vec{J}_{\vec{n}}\: ,
\end{equation}

\noindent where $\vec{J}_{\vec{R}}$ and $\vec{J}_{\vec{n}}$ are the corresponding diffusion currents. Following the steps indicated before we then obtain the entropy production which has the usual form of a sum of fluxes-force pairs

\begin{equation}\label{eq:b12}
\sigma = -\frac{1}{T}\int d\vec{R}\, d\vec{n}\,\vec{J}_{\vec{R}}\cdot\nabla_{\vec{R}}\mu -\frac{1}{T}\int d\vec{R}\,d\vec{n}\, \vec{J}_{\vec{n}}\cdot\nabla_{\vec{n}}\mu\; .
\end{equation}

\noindent From this equation one then derives the linear laws

\begin{eqnarray}\label{eq:b13}
\vec{J}_{\vec{R}} &=& -\frac{1}{T}\underline{L}_{\vec{R},\vec{R}}\cdot\nabla_{\vec{R}}\mu - \frac{1}{T}\underline{L}_{\vec{R},\vec{n}}\cdot\nabla_{\vec{n}}\mu\\ \nonumber
\vec{J}_{\vec{n}} &=& -\frac{1}{T}\underline{L}_{\vec{n},\vec{n}}\cdot\nabla_{\vec{n}}\mu - \frac{1}{T}\underline{L}_{\vec{n},\vec{R}}\cdot\nabla_{\vec{R}}\mu\; ,
\end{eqnarray}

\noindent where $\underline{L}_{\alpha\beta}$, $\alpha, \beta = (\vec{R},\vec{n})$, are matrices of phenomenological coefficients, written in local form. Alternatively, by using the expression for the chemical potential, $\mu(\vec{n},\vec{R},t) = \frac{kT}{N}\ln P(\vec{n},\vec{R},t)$ (for simplicity sake we do not consider the action of any field of forces), these equations can be rewritten as

\begin{eqnarray}\label{eq:b14}
\vec{J}_{\vec{R}} &=& -\underline{M}_{\vec{R},\vec{R}}\cdot\nabla_{\vec{R}}P - \underline{M}_{\vec{R},\vec{n}}\cdot\nabla_{\vec{n}}P\\ \nonumber
\vec{J}_{\vec{n}} &=& -\underline{M}_{\vec{n},\vec{n}}\cdot\nabla_{\vec{n}}P - \underline{M}_{\vec{n},\vec{R}}\cdot\nabla_{\vec{R}}P\; ,
\end{eqnarray}

\noindent where use has been made of the relation among the mobilities and the phenomenological coefficients. According to Onsager's symmetry principle, the phenomenological coefficients satisfy the relations

\begin{equation}\label{eq:b15}
\underline{M}_{\vec{R},\vec{n}} = \underline{M}_{\vec{n},\vec{R}}^\dagger\; .
\end{equation}

\noindent By substituting the linear equations (\ref{eq:b14}) into the continuity equation (\ref{eq:b11}) we then obtain the kinetic equation

\[\frac{\partial P}{\partial t} = - \nabla_{\vec{R}}\cdot\left(\underline{M}_{\vec{R},\vec{R}}\cdot\nabla_{\vec{R}}P - \underline{M}_{\vec{R},\vec{n}}\cdot\nabla_{\vec{n}}P\right)\]
\begin{equation}\label{eq:b16}
-\nabla_{\vec{n}}\cdot\left(\underline{M}_{\vec{n},\vec{n}}\cdot\nabla_{\vec{n}}P - \underline{M}_{\vec{n},\vec{R}}\cdot\nabla_{\vec{R}}P\right)\; .
\end{equation}

Up to lowest order in the parameter $b/L$, it is known \cite{kn:doi} that the mobility matrices can be written as

\[\underline{M}_{\vec{R},\vec{R}} = \mu_{\parallel}^t \vec{n}\vec{n} + \mu_{\perp}^t (\underline{\bf 1} - \vec{n}\vec{n})\]
\[\underline{M}_{\vec{n},\vec{n}} = \mu^r (\underline{\bf 1} - \vec{n}\vec{n})\]
\begin{equation}\label{eq:b17}
\underline{M}_{\vec{n},\vec{R}} = \underline{M}_{\vec{R},\vec{n}} = 0\; ,
\end{equation}

\noindent where $\mu_{\parallel,\perp}^t$ and $\mu^r$ are the translational and rotational scalar mobilities, and $\underline{\bf 1}$ is the unit tensor. Insertion of Eqs. (\ref{eq:b17}) into (\ref{eq:b16}) then leads to the Smoluchowski equation

\begin{equation}\label{eq:b18}
\frac{\partial P}{\partial t} = \nabla_{\vec{R}}\cdot\left\{\left[D_{\parallel}\vec{n}\vec{n} + D_{\perp}(\underline{\bf 1} - \vec{n}\vec{n})\right]
\cdot\nabla_{\vec{R}} P\right\} + D_r \vec{\cal R}^2 P\; ,
\end{equation}

\noindent where we have introduced the diffusion coefficients $D_{\parallel} = kT\mu_{\parallel}^t$, $D_{\perp} = kT\mu_{\perp}^t$, and $D_r = kT\mu^r$ and $\vec{\cal R} = \vec{n}\times \partial/\partial\vec{n}$ is the rotational operator.

\section{semiflexible polymers}

Unlike flexible polymers, semiflexible polymers characterize by the presence of some degree of stiffness that introduces a persistence length \cite{kn:freed}. The semiflexible polymer may be described by a continuous curve $\vec{r}(l)$, with $l$ ($0\leq l\leq L$) being the contour length along the chain. To derive the Fokker-Planck equation, we will introduce  the continuity equation

\begin{equation}\label{eq:d1}
\frac{\partial }{\partial t}P(\{\vec{r}(l)\},t) = -\int dl^\prime \frac{\delta}{\delta \vec{r}(l^\prime)}\cdot\vec{J}(l^\prime,t) \; .
\end{equation}

\noindent Here $N^{-1}P(\{\vec{r}(l)\},t)$ represents the probability density for the configuration $\vec{r}(l)$ at time t, $\vec{J}(l,t)$ is the diffusion current in the configuration space. 

The rate of change of entropy can be obtained from the corresponding Gibbs equation, and reads

\begin{equation}\label{eq:d2}
\frac{\partial S}{\partial t} = -\frac{1}{T}\int\delta \vec{r}(l)\mu(\{\vec{r}(l)\},t)\frac{\partial }{\partial t}P(\{\vec{r}(l)\},t) \; .
\end{equation}

\noindent Here, the chemical potential is given by

\begin{equation}\label{eq:d3}
\mu(\{\vec{r}(l)\},t) = \frac{kT}{N}\ln P(\{\vec{r}(l)\},t) +\frac{1}{N} U(\{\vec{r}(l)\})\; ,
\end{equation}

\noindent where the potential $U(\{\vec{r}(l)\})$ consists of  entropic  and elastic energy of bending contributions \cite{kn:freed} 

\begin{equation}\label{eq:d4}
U(\{\vec{r}(l)\})=\frac{3}{2l_0}kT\int_0^L \left(\frac{\partial\vec{r}(l)}{\partial l}\right)^2 dl + \frac{\epsilon}{2}\int_0^L \left(\frac{\partial^2\vec{r}(l)}{\partial l^2}\right)^2 dl
\end{equation}

\noindent with $l_0$ being the effective segment length, and $\epsilon$  the bending force constant.

By making use of Eq. (\ref{eq:d1}), and after partial integration we obtain the entropy production related to the diffusion process in the configurational space

\begin{equation}\label{eq:d5} 
\sigma = -\frac{1}{T}\int\delta \vec{r}(l)\,\int dl\, \vec{J}\cdot\frac{\delta\mu}{\delta \vec{r}(l)}\; .
\end{equation}

\noindent From the entropy production we derive the linear laws

\begin{equation}\label{eq:d6} 
\vec{J}(l,t) = -\frac{1}{T}\int dl^\prime\,\underline{L}(l,l^\prime)\cdot\frac{\delta\mu}{\delta \vec{r}(l^\prime)}\; ,
\end{equation}

\noindent where we have taken into account that now locality is not fulfilled. The matrices of Onsager coefficients $\underline{L}(l,l^\prime)$ satisfy de Onsager's relations

\begin{equation}\label{eq:d7}
\underline{L}(l,l^\prime) = \underline{L}(l^\prime,l)^\dagger\; .
\end{equation}

\noindent Using Eqs. (\ref{eq:d6}) with (\ref{eq:d3}) in the continuity equation (\ref{eq:d1}),  we obtain the functional Fokker-Planck equation

\begin{equation}\label{eq:d8}
\frac{\partial P}{\partial t} = \int dl \int dl^\prime \frac{\delta}{\delta \vec{r}(l)}\cdot\underline{M}(l,l^\prime)\cdot\left\{kT\frac{\delta}{\delta \vec{r}(l^\prime)}P + P\frac{\delta}{\delta \vec{r}(l^\prime)}U(\{\vec{r}(l^\prime)\}) \right\}
\end{equation}

\noindent describing the evolution of the probability density for the configurations of the polymer.

\section{many chain systems}

In this section, we will indicate how the formalism developed in previous sections for dilute polymer solutions  also applies to the case of higher concentrations. We will analyze the case of flexible polymers and consider as the internal coordinate the vector of the configuration space $\underline{\gamma}\equiv\{\vec{R}_\nu^a\}$, with $\vec{R}_\nu^a$ being the position vector of the $\nu$-th segment pertaining to the 
$a$-th chain. The probability density in the configuration space is then  $N^{-1}P(\{\vec{R}_\nu^a\},t)$ where now $N$ must be understood as the total number of segments in suspension. As we did in section 3, the starting point in our analysis is the partial Gibbs equation for the polymer phase. This equation together with the continuity equation

\begin{equation}\label{eq:c1}
\frac{\partial P}{\partial t} = - \sum_{a,\nu}\nabla_{\vec{R}_\nu^a}\cdot\vec{J}_{\nu}^a
\end{equation}

\noindent allows us to express the entropy production as

\begin{equation}\label{eq:c2}
\sigma = -\frac{1}{T}\int\sum_{a,\nu}\vec{J}_{\nu}^a\cdot\nabla_{\vec{R}_\nu^a}\mu\;  d\{\vec{R}_\nu^a\}\; ,
\end{equation}

\noindent where $\vec{J}_{\nu}^a$ are currents in the configurational space spanned by the values of the coordinates $\{\vec{R}_\nu^a\}$. Moreover, the chemical potential consists also of ideal and non-ideal contributions and is given by

\begin{equation}\label{eq:c3}
\mu(\{\vec{R}_\nu^a\},t) = \frac{kT}{N}\ln P(\{\vec{R}_\nu^a\},t) +\frac{1}{N} U(\{\vec{R}_\nu^a\})\; .
\end{equation}

The linear laws are then written as

\begin{equation}\label{eq:c4}
\vec{J}_{\nu}^a = -\frac{1}{T}\sum_{b,\eta}\underline{L}_{\nu\eta}^{ab}\cdot\nabla_{\vec{R}_\eta^b}\mu\; ,
\end{equation}

\noindent where $\underline{L}_{\nu\eta}^{ab}$ are matrices of phenomenological coefficients relating the current $\vec{J}_{\nu}^a$, associated with the internal coordinate $\{\vec{R}_\nu^a\}$, with its conjugated thermodynamic force $\nabla_{\vec{R}_\eta^b}\mu$. These matrices satisfies the Onsager relations

\begin{equation}\label{eq:c5}
\underline{L}_{\nu\eta}^{ab} = {\underline{L}_{\eta\nu}^{ba}}^\dagger\; .
\end{equation}

\noindent Substitution of Eq. (\ref{eq:c4}) in Eq. (\ref{eq:c1}) then leads to the kinetic equation

\begin{equation}\label{eq:c6}
\frac{\partial P}{\partial t} = - \sum_{a,b,\nu\,\eta}\nabla_{\vec{R}_\nu^a}\cdot\underline{M}_{\nu\eta}^{ab}\cdot\left\{ kT\nabla_{\vec{R}_\eta^b}P + P\nabla_{\vec{R}_\eta^b}U\right\}\; .
\end{equation}

Alternatively,  we could now proceed to describe the distribution of the polymer segments in terms of the local segment density $c(\vec{r}) = \sum_{a,\nu}\delta(\vec{r}-\vec{R}_\nu^a )$\cite{kn:doi}. This description, which is useful when performing ensemble averages of quantities depending on the local segment density, would then imply the introduction of the probability density $P(\{c(\vec{k})\},t)$ \cite{kn:rubi} in the Gibbs equation in order to specify the local state of the polymer solution. Proceeding along the lines indicated in previous cases, we obtain the entropy production 

\begin{equation}\label{eq:c7}
\sigma = -\frac{1}{T}\int \delta c(\vec{k})\int d\vec{k} J(\{c(\vec{k})\},t;\vec{k})\frac{\delta}{\delta c(\vec{k})}\mu(\{c(\vec{k})\},t)\; ,
\end{equation}

\noindent where $J(\{c(\vec{k})\},t;\vec{k})$ is the corresponding current and the chemical potential is given by

\begin{equation}\label{eq:c8}
\mu(\{c(\vec{k})\}) = \frac{kT}{N}\ln\rho(\{c(\vec{k})\}) + \frac{1}{N}U(\{c(\vec{k})\}\; .
\end{equation}

\noindent In Eq. (\ref{eq:c7}) we have introduced functional integrals and derivatives because the continuum character of the index $\vec{k}$, with $k_z \geq 0$. The linear relations are now written in non-local form as

\begin{equation}\label{eq:c9}
J(\{c(\vec{k})\},t;\vec{k}) = -\frac{1}{T}\int d\vec{k}^\prime L(\vec{k},\vec{k}^\prime)\frac{\delta}{\delta c(\vec{k}^\prime)}\mu(\{c(\vec{k}^\prime),t)\; .
\end{equation}

\noindent Following the procedure indicated previously we finally arrive at the Fokker-Planck equation  

\begin{equation}\label{eq:c10}
\frac{\partial}{\partial t}P(\{c(\vec{k})\}) = \int d\vec{k} \int d\vec{k}^\prime \frac{\delta}{\delta c(\vec{k})}M(\vec{k},\vec{k}^\prime)\left\{kT\frac{\delta}{\delta c(\vec{k}^\prime)}P(\{c(\vec{k^\prime})\}) + P\frac{\delta}{\delta c(\vec{k}^\prime)}U(\{c(\vec{k^\prime})\}) \right\}\; ,
\end{equation}

\noindent where for the sake of simplicity we have not considered the effects of an external flow. For the case in which the matrix $L(\vec{k},\vec{k}^\prime)$ is local in Fourier space due to the preaveraging approximation, this equation agrees with the one obtained in Ref. \cite{kn:doi} from the Langevin equation for the density. In the method we have presented, the kinetic equation comes from the analysis of the dissipation in phase space, according to the principles of nonequilibrium thermodynamics.

\section{conclusions}

In this paper, we have applied the method of mesoscopic nonequilibrium thermodynamics to derive the kinetic equations governing the dynamics of macromolecules in solution. The Fokker-Planck equations can be obtained from the analysis of the underlying driven diffusion process occurring in the space of configurations of the polymers.

The intrinsic nature of the polymer defines the set of degrees of freedom necessary for its complete dynamic description. One then formulates the continuity equation for the probability density in the configuration space which introduces an unspecified current in that space. The determination of that current is carried out through nonequilibrium thermodynamics giving rise to the Fokker-Planck equation. The simplicity of the procedure contrasts with the more elaborated method of deriving the Fokker-Planck equation from the corresponding Langevin equation, becoming more cumbersome for complex systems involving many degrees of freedom.

The formalism we have proposed can be applied to polymers of different nature in the presence of external gradients \cite{kn:senger} or forces and for arbitrary values of the concentration. It may then constitute a useful tool in the theory of polymer dynamics.

\acknowledgments

We want to acknowledge D. Reguera for his comments. This work has been supported by DGICYT of the Spanish Government under grant PB98-1258,

\end{document}